\documentstyle[epsfig,multicol,aps,prl]{revtex}

\begin{document}
\draft 

\title{Invariance of Charge of Laughlin Quasiparticles}

\author{V. J. Goldman$^{a}$, I. Karakurt$^{a}$, Jun Liu$^{b}$, and A. Zaslavsky$^{b}$}
\address{$^{a}$Department of Physics, State University of New York, Stony Brook, NY 11794-3800\\
$^{b}$Department of Electrical Engineering, Brown University, Providence, RI 02912\\}

\date{September 6, 2000}

\maketitle

\begin{abstract}
A Quantum Antidot electrometer has been used in the first direct observation of the \mbox{fractionally} quantized electric charge. In this paper we report experiments performed on the integer $i = 1, 2$ and \mbox{fractional} $f = 1/3$ quantum Hall plateaus extending over a filling factor range of at least 27\%. We find the charge of the Laughlin quasiparticles to be invariantly $e/3$, with standard deviation of $1.2\%$ and absolute accuracy of $4\%$, independent of filling, tunneling current, and temperature.

\end{abstract}

\pacs{PACS numbers: 73.40.Hm, 73.40.Gk, 71.10.Pm}

\begin{multicols}{2}

The most fundamental aspect of the quantum Hall effect (QHE) is the constancy of the Hall conductance over a finite range of the filling factor $\nu$.\cite{vonK,Tsui82,Books} Indeed, this property defines the phenomenon of QHE, and the quantized value of the Hall conductance $\sigma_{xy}$ of a particular QH state in units of $e^2/h$ is a principal quantum number of that QH state $i$ or $f$, called ``exact filling''. Specifically, the electric charge of the quasiparticles is expected to be determined by the relevant quantum numbers, including $i$ or $f$, and thus not expected to vary on a QH plateau as $\nu$ is varied from the exact filling.\cite{Books,Lgauge,HalpEdge} 

On a QH plateau the charge of quasiparticles localized in the interior of a two dimensional electron system (2DES) is well defined.\cite{Wen} In the case of the integer QH plateau at $\nu \approx i$ the quasielectrons are simply electrons in the Landau level $i+1$, and the quasiholes are the holes in the $i^{th}$ level. It is easy to understand the properties of FQHE quasiparticles using composite fermions.\cite{Jain89,JG} In the case of the FQH plateau at $f =\frac{i}{2pi+1}$ quasielectrons are composite fermions (an electron binding $2p$ vortices) in the ``Landau level" $i+1$ of composite fermions, and the quasiholes are the holes in the $i^{th}$ level. It has been predicted theoretically that the electric charge of these quasiparticles is $q=\frac{e}{2pi+1}$, positive for the quasiholes and negative for quasielectrons. \cite{Laughlin83,Books} This fascinating fractional quantization of electric charge is a fundamental property of the strongly correlated FQH fluid.

Although six years have passed since the first direct observation of $\frac{1}{3}e$ particles in quantum antidot (QAD) electrometer experiments\cite{SciCharge} one crucial aspect of theory remained untested: the invariance of charge at $\nu $ far from exact filling. In this paper we report experiments performed on the integer $i = 1, 2$ and fractional $f=\frac{1}{3}$ quantum Hall plateaus which extend over a filling factor $\nu$ range of 27$\%$ to 45$\%$. The charge of the QAD-bound quasiparticles has been measured to be constant, independent of $\nu$ over the entire plateau extent, with relative accuracy of $\pm 1.2\%$ and absolute accuracy of $4\%$. In addition, we observe no variation of the quasiparticle charge upon variation of temperature or applied current, in the experimentally accessible range.

The QAD electrometer is illustrated in Fig. 1.\cite{SciCharge,SurfSciE} The antidot is defined lithographically in a constriction between two front gates in a 2DES. The antidot and the front gates create depletion potential hills in the 2DES plane and, in quantizing magnetic field $B$, the QHE edge channels are formed following equipotentials where the electron density $n$ is such that $\nu = en/hB$ is equal to integer $i$ or fractional $f$ exact filling. The edge channels on the periphery of the 2DES have a continuous energy   
\begin{figure}
\centerline{\epsfig{file=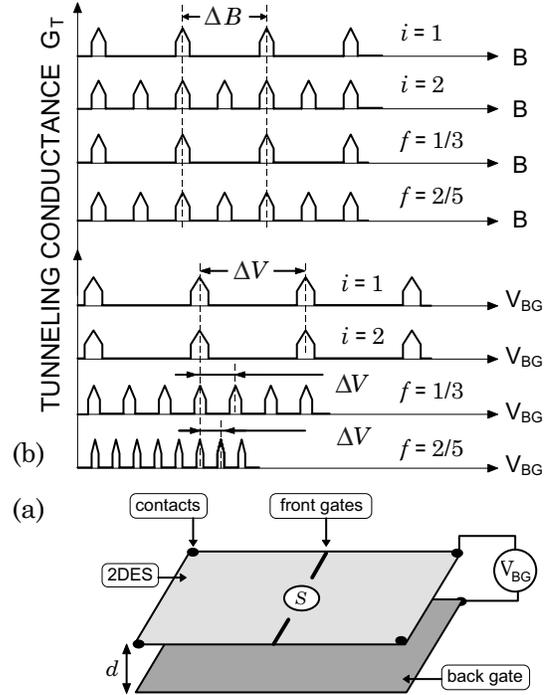, width=7.1cm}}
\vspace{0.2cm} 
\caption{(a) Schematic of the quantum antidot electrometer. (b) Idealized summary of experimental observations. There is one $G_T$ peak on the $i=1$ and $f=\frac{1}{3}$ plateaus, and two peaks on the $i=2$ and $f=\frac{2}{5}$ plateaus observed per period $\Delta B$, with the back gate voltage held constant $V_{BG}=0$.  $\Delta B$ gives the size of the QAD: $S=\phi_0/\Delta B$. The same $G_T$ peaks are also observed when a small perpendicular electric field $V_{BG}/d$ is applied to 2DES by biasing the back gate. The period $\Delta V_{BG}$ then directly gives the charge $q$ of the QAD-bound particles, in Coulombs, via $q=(\epsilon \epsilon_0 \phi_0/d)(\Delta V_{BG}/\Delta B)$. }
\label{Fig. 1}
\end{figure}
\noindent 
spectrum, while the particle states of the edge channel circling the antidot are quantized by the Aharonov-Bohm condition that the state $\psi_m$ with quantum number $m$, in each Landau level,  encloses $m\phi_0=m\frac{h}{e}$ magnetic flux.  In other words, the semiclassical area of the state $\psi_m$ is $S_m = m\phi_0/B$. The electrometer application is made possible by a large, global ``back gate'' on the other side of the GaAs sample of thickness $d$. This gate forms a parallel plate capacitor with the 2D electrons. 

When the constriction is on a quantum Hall plateau, particles can tunnel resonantly via the QAD-bound states giving rise to quasiperiodic tunneling conductance $G_T$ peaks, see Fig. 1.\cite{JainKivelson}  A peak in $G_T$ occurs when a QAD-bound state crosses the chemical potential and thus marks the change of QAD occupation by one particle.  The experimental fact that the $G_T$ peaks are observed implies that the charge induced in the QAD is quantized, that it comes in discrete particles occupying the antidot-bound states; the $G_T$ peaks mark the change of the population of the QAD by one particle per peak.  Measuring $G_T$ as a function of $B$ gives the area of the QAD-bound state through which the tunneling occurs, $S=\phi_0/\Delta B$, where $\Delta B$ is the quasiperiod in magnetic field.  On the $\nu \approx i$ plateau there are $i$ peaks per $\Delta B$ because $i$ Landau levels are occupied. The above discussion is easy to generalize for $f =\frac{i}{2pi+1}$ FQH plateaus by considering ``Landau levels'' of composite fermions.

The charge of the QAD-bound particles is then determined directly from the separation of the same $G_T$ peaks as a function of the back gate voltage $V_{BG}$.\cite{impurities} The back gate produces uniform electric field $E_{\perp} =V_{BG}/d$ which induces a small change of $\epsilon \epsilon_0 E_{\perp}$ in the 2DES charge density.\cite{sigmaxx} Classical electrostatics states that the charge induced in the QAD is exactly equal to the 2D charge density induced {\it far} from the QAD times the area of the QAD, $\epsilon \epsilon_0 E_{\perp}/S$.\cite{LLECM,FrontGate}  Thus, the charge of one particle $q$ is directly given by the electric field needed to attract one more particle in the area $S$: $q=\epsilon \epsilon_0 S\Delta V_{BG}/d$, where $\Delta V_{BG}$ is the change of the global gate voltage between two consecutive conductance peaks. \cite{SciCharge,SurfSciE}

We use low disorder GaAs heterojunction material where 2DES (density $1{\small\times}10^{11}$ cm$^{-2}$ and mobility $2{\small\times} 10^{6}$ cm$^{2}/V s$) is prepared by exposure to red light at 4.2 K. The antidot-in-a-constriction geometry (somewhat different from that of Refs.\cite{SciCharge,SurfSciE}) was defined by electron beam lithography on a pre-etched mesa with Ohmic contacts. After $\approx$150 nm chemical etching Au/Ti gate metalization was deposited in the etched trenches.  Samples were mounted on sapphire substrates with In metal which serves as the global back gate.  All data presented in this paper were taken at 12 mK bath temperature. Extensive cold filtering cuts the electromagnetic background incident on the sample to 5${\small\times}10^{-17}$ W, which allows us to achieve a record low effective electron temperature of 18 mK reported for a mesoscopic sample.\cite{Maasilta97}

Figures 2 - 4 show the directly measured four-terminal $R_{xx}$ {\it vs.} $B$ data for three QAD plateaus: $i=2$, $i=1$ and $f=\frac{1}{3}$. We use $\nu $ to denote the filling factor in the constriction region. The front gates are biased negatively in order to bring the edges closer to the antidot to increase the amplitude of the tunneling peaks to a measurable level.  This results in $\nu$ being smaller than $\nu_B$ in the rest of the sample (``the bulk'').  A QHE sample with two $\nu_B$ regions separated by a lower $\nu$ region, if no tunneling occurs, has $R_{xx}=R_L\approx R_{xy}(\nu) - R_{xy}(\nu_B)$. The equality is exact if both $\nu$ and $\nu_B$ are on a plateau, where the Hall  
\end{multicols}
\noindent
\begin{figure}
\centerline{\epsfig{file=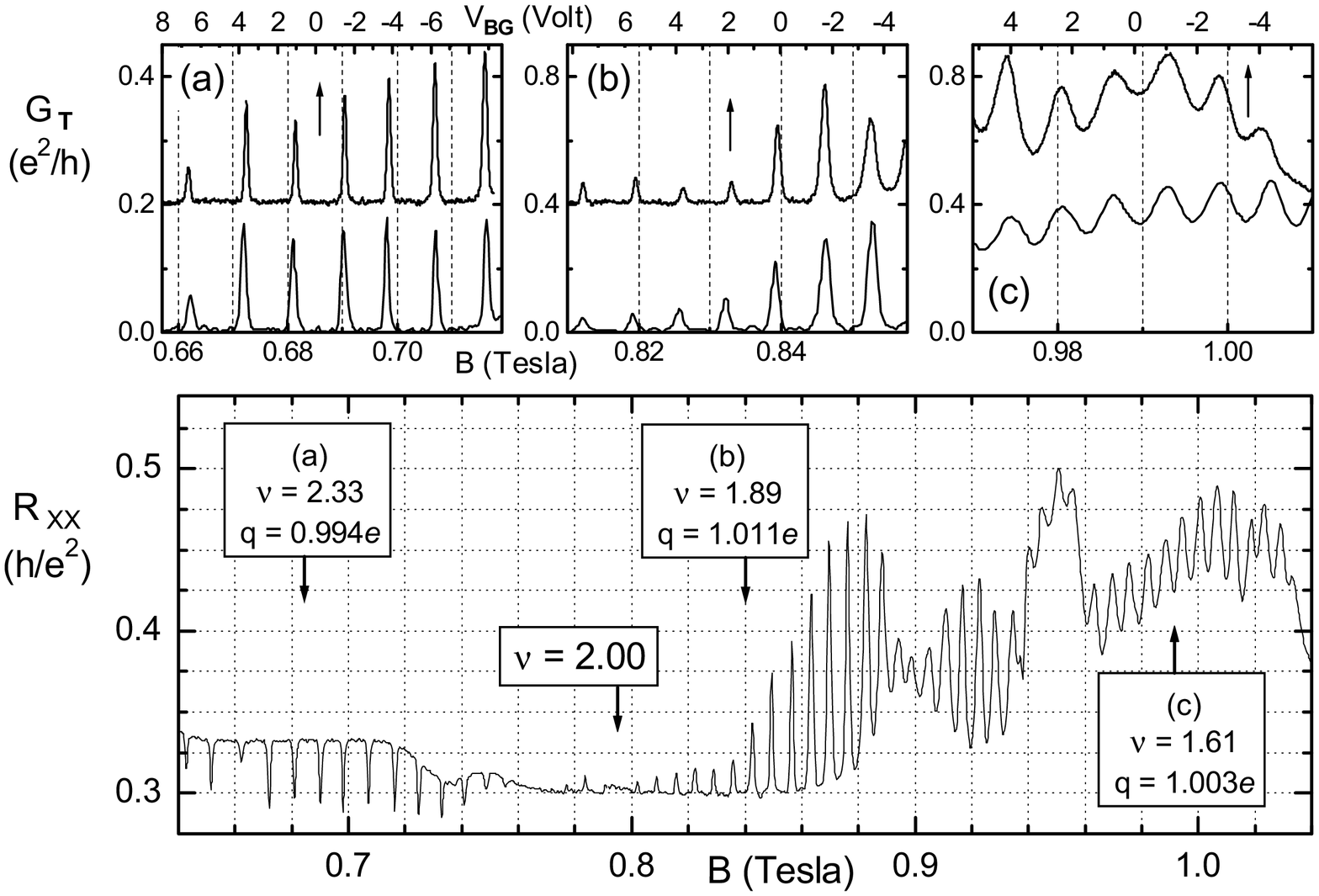, width=11.4cm}}  
\vspace{0.1cm} 
\caption{$R_{xx}$ {\em vs.} magnetic field $B$ for the antidot filling $\nu$ on the $i=2$ plateau. The upper panels (a) - (c) give the tunneling conductance measured both as a function of $B$ (back gate voltage is held constant $V_{BG}=0$), and as a function of $V_{BG}$ ($B$ is held constant, shown by arrows in the lower panel). The $G_T$ {\em vs.} $V_{BG}$ curves in panels (a) - (c) are offset vertically. }
\label{Fig. 2}
\end{figure}

\begin{figure}
\centerline{\epsfig{file=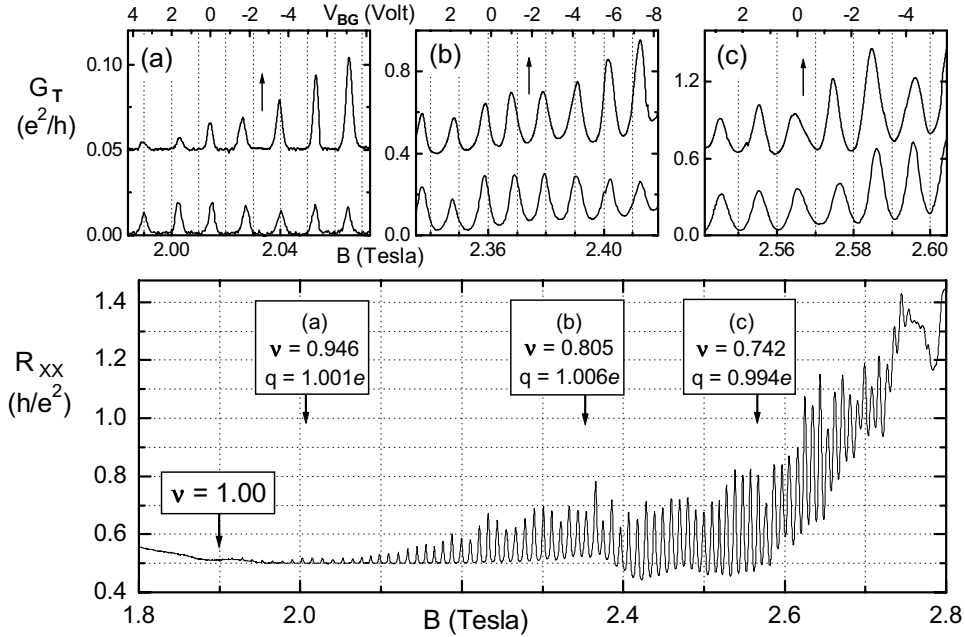, width=12.7cm}}  
\vspace{0.1cm} 
\caption{$R_{xx}$ {\em vs.} magnetic field $B$ for the antidot filling $\nu$ on the $i=1$ plateau. The upper panels (a) - (c) give the tunneling conductance measured both as a function of $B$ (back gate voltage $V_{BG}=0$), and as a function of $V_{BG}$ ($B$ is held constant).}
\label{Fig. 3}
\end{figure}
\begin{multicols}{2}
\noindent
resistances of all regions aquire quantized values. Thus, several $R_L$ plateaus (neglecting tunneling peaks) are seen in Figs. 2 - 4. The tunneling peaks are superimposed on the smooth $R_L$ background, and we calculate $G_T$ as described previously.\cite{SciCharge,Maasilta97} In some data (Figs. 2 and 4) we observe both $R_{xx}$ peaks for $\nu <i$ (``back scattering'') and dips for $\nu >i$ (``forward scattering'').\cite{JainKivelson,Ford94} Details of this behavior will be presented elsewhere. 

At several $\nu$ on each plateau we took high resolution $B$-sweeps at $V_{BG}=0$, and, having put the superconducting magnet in the persistent current mode to fix $B$, we took corresponding sweeps of $V_{BG}$.  Representative $G_T$ {\it vs.} $B$ and $V_{BG}$ data are shown in the upper panels of Figs. 2 - 4.  Note that the negative $V_{BG}$ axis direction corresponds to the increasing $B$. This is so because increasing $B$ results in relative depopulation of QAD independent of the sign of the charge of particles: $\psi_m$ move closer to the center of the QAD, that is, more states move above  
\end{multicols}
\begin{figure}
\centerline{\epsfig{file=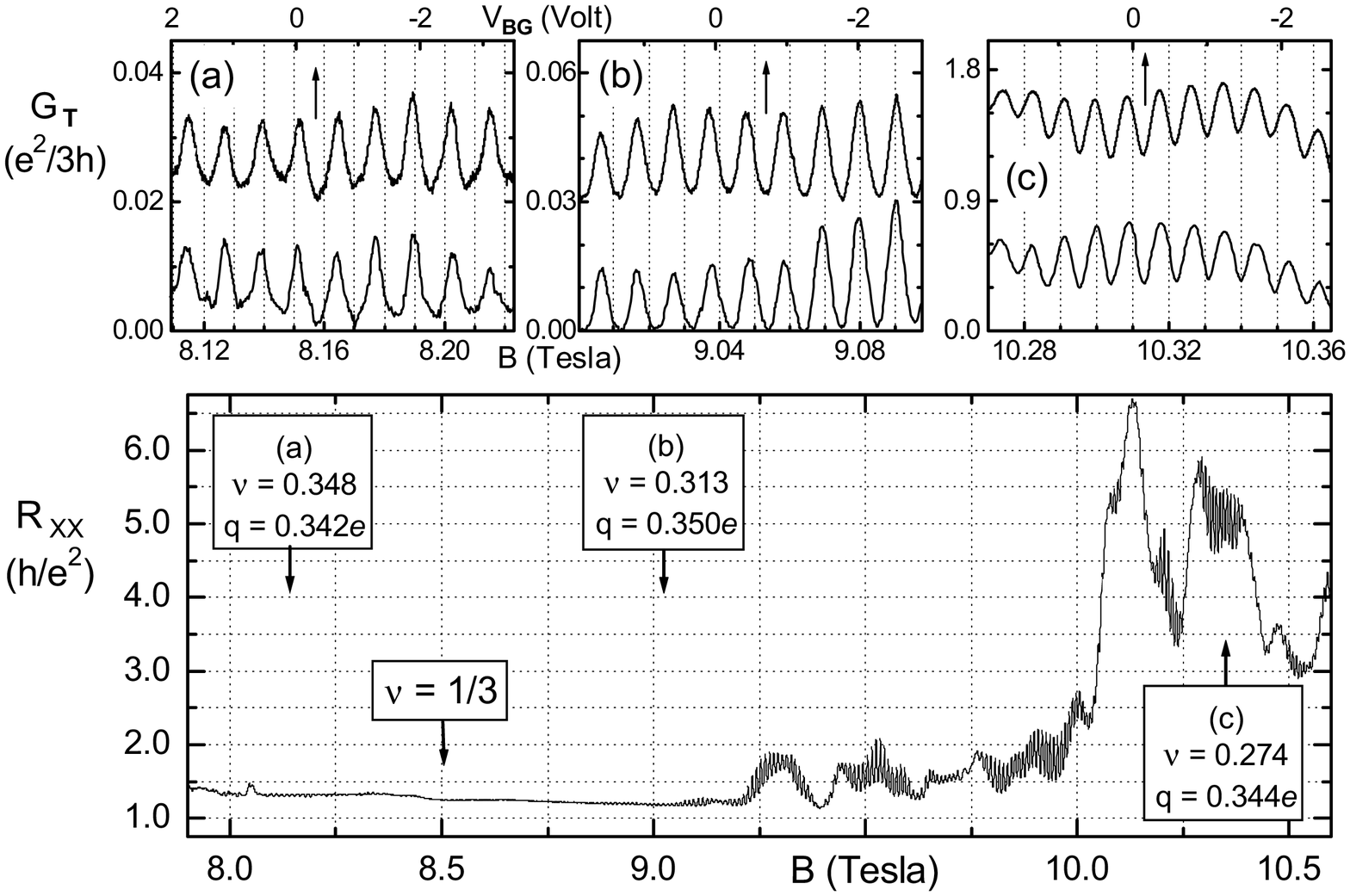, width=12.7cm}}  
\vspace{0.1cm} 
\caption{$R_{xx}$ {\em vs.} magnetic field $B$ for the antidot filling $\nu$ on the $f=\frac{1}{3}$ plateau. The upper panels (a) - (c) give the tunneling conductance measured both as a function of $B$ (back gate voltage $V_{BG}=0$), and as a function of $V_{BG}$ ($B$ is held constant).}
\label{Fig. 4}
\end{figure}
\begin{multicols}{2}

\begin{figure}
\centerline{\epsfig{file=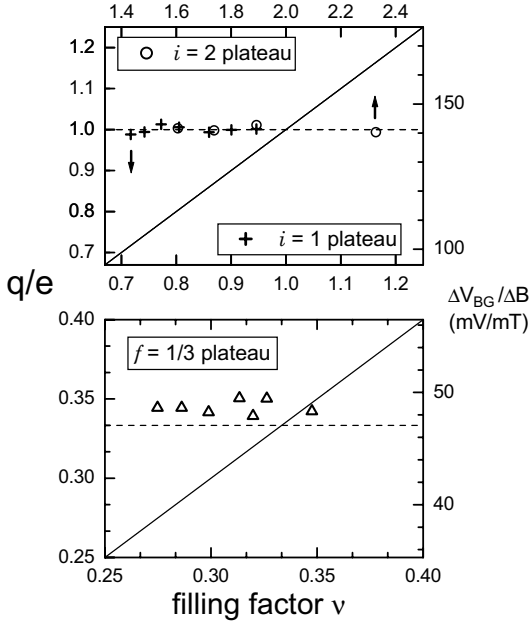, width=7cm}}
\vspace{0.1cm} 
\caption{Summary of the measured QAD-bound quasiparticle charge $q$ in units of $e$. The horizontal dashed lines give the values $e$ and $e/3$. The solid lines have the unit slope.  }
\label{Fig. 5}
\end{figure}
\noindent
 the chemical potential and become unoccupied.  Negative $V_{BG}$ also depopulates the QAD if QAD-bound particles have {\em negative} charge.  Thus the QAD electrometer measures not only the magnitude of $q$, but also its sign.

The magnitude of the charge $q$ of the QAD-bound particles is then determined from 
\begin{equation} 
q=\frac{\epsilon \epsilon_0 \phi_0}{d}\cdot \frac{\Delta V_{BG}}{\Delta B} $~~~in Coulombs,$
\label{Eq. 1}
\end{equation} 
\noindent
using the low-temperature GaAs dielectric constant $\epsilon=12.74$\cite{epsilonGaAs} and the measured thickness of the sample $d \approx 0.430$ mm.  The average of the eleven values obtained for the $i=1, 2$ plateaus $\left< q\right>_{\small {\rm integer}}=0.9651e$ is off by 3.5\%, more than the standard deviation of $0.0070e$ for the combined data (similar to the results from another electrometer device reported in Refs. \cite{SciCharge,SurfSciE}).  We then {\it normalize} values of $q$ by setting $\left< q\right>_{\small {\rm integer}} = e$.  Thus determined $q$ are shown for (a) - (c) data in Figs. 2 - 4, and summarized in Fig. 5. The striking feature of the data of Fig. 5 is that the values of $q$ are constant to a relative accuracy of at least $\pm1.2\%$ throughout the plateau regions where it was possible to measure the particle charge. The range of $\nu$ is about 45\% for the $i=2$ plateau, 27\% for the $i=1$ plateau (the combined normalized $\nu/i$ range is 57\%), and also 27\% for the $f=\frac{1}{3}$ plateau.  

We also note that the tunneling current $I_t \approx I G_T/ \sigma_{xy}$ is proportional to $G_T$ and thus varies much for peaks of different amplitude. Here $I$ is the applied current used to measure $R_{xx}$; the $f=\frac{1}{3}$ plateau data of Fig. 4 was taken with $I=50$ pA, and we have measured $\Delta V_{BG}/\Delta B$ with $I$ up to 1 nA, which, combined with the variation in $G_T$ gives the range of $5{\small \times}10^{-13}$ A $\leq I_t\leq 2{\small \times}10^{-10}$ A.  Furthermore, $\Delta V_{BG}/\Delta B$ has been measured in the temperature range 12 mK $\leq T\leq$ 70 mK.\cite{Maasilta97}  Under these conditions we observe no change in the value of $q=e/3$ within our experimental accuracy of a few percent. 
Thus we conclude that the charge of the quasiparticles is indeed a well defined quantum number characterizing a particular QHE state.

We are grateful to J. K. Jain for discussions. This work was supported in part by the NSF under Grant No. DMR9986688. The work at Brown was supported by an NSF Career Award DMR9702725 and the NSF MRSEC Center DMR9632524.

\vspace{-0.1 cm}
\bibliographystyle{prsty}

\begin{references}

\bibitem{vonK} K. von Klitzing, G. Dorda, and M. Pepper, Phys. Rev. Lett. {\bf 45}, 494 (1980).
\bibitem{Tsui82} D. C. Tsui, H. L. Stormer, and A. C. Gossard, Phys. Rev. Lett. {\bf 48}, 1559 (1982).
\bibitem{Books} For reviews see: {\em The Quantum Hall Effect}, 2nd Ed. (Ed. by R. E. Prange and S. M. Girvin) Springer, NY (1990); {\em Perspectives in Quantum Hall Effects}, (Ed. by S. Das Sarma and A. Pinczuk) Wiley, NY (1997); S. M. Girvin, {\em The Quantum Hall Effect}, Les Houches Lecture Notes (1998); V. J. Goldman, Physica B {\bf 280}, 372 (2000).

\bibitem{Lgauge} R. B. Laughlin, Phys. Rev. B {\bf 23}, 5632 (1981).
\bibitem{HalpEdge} B. I. Halperin, Phys. Rev. B {\bf 25}, 2185 (1982).
\bibitem{Wen} Excitations of arbitrary charge can be created in edges, where there is no gap; e.g., X.-G. Wen, Intl. J. Mod. Phys. B {\bf 6}, 1711 (1992). A weakly coupled QAD serves as a ``charge filter'' for interedge tunneling.

\bibitem{Jain89} J. K. Jain, Phys. Rev. Lett. {\bf 63}, 199 (1989).
\bibitem{JG} J. K. Jain and V. J. Goldman, Phys. Rev. B {\bf 45}, 1255 (1992).
\bibitem{Laughlin83} R. B. Laughlin, Phys. Rev. Lett. {\bf 50}, 1395 (1983).

\bibitem{SciCharge} V. J. Goldman and B. Su, Science {\bf 267}, 1010 (1995).
\bibitem{SurfSciE} V. J. Goldman, Surf. Sci. {\bf 361/362}, 1 (1996); Physica E {\bf 1}, 15 (1997).

\bibitem{JainKivelson} J. K. Jain and S. A. Kivelson, Phys. Rev. Lett. {\bf 60}, 1542 (1988).
\bibitem{impurities} Unintentional impurities near the antidot can change their charge state as $B$ or $V_{BG}$ are swept. Such events, often reproducible, cause ``phase slips'' of the $G_T$ peaks; we exclude such data from the charge analysis.

\bibitem{sigmaxx} Even on a QHE plateau it is possible to change 2DES density in the ``incompressible'' interior with a gate so long as it is done slowly on the time scale of $\tau=RC\approx \epsilon \epsilon_0 L^2/(d\sigma_{xx})$, where $L$ is the  size of 2DES, because of finite diagonal conductivity $\sigma_{xx}$ at a finite temperature.
\bibitem{LLECM} L. D. Landau and E. M. Lifshitz, {\em Electrodynamics of Continuous Media}, Chap. 1, Sect. 3.
\bibitem{FrontGate} A small front gate that {\em defines} the antidot (as in J. Franklin {\em et al.}, Surf. Sci. {\bf 361/362}, 17 (1996)) clearly does not produce a uniform $E_{\perp}$.

\bibitem{Maasilta97} I. J. Maasilta and V. J. Goldman, Phys. Rev. B {\bf 55}, 4081 (1997); Phys. Rev. Lett. {\bf 84}, 1776 (2000).
\bibitem{Ford94} C. J. B. Ford {\em et al.}, Phys. Rev. B {\bf 49}, 17456 (1994).
\bibitem{epsilonGaAs} G. A. Samara, Phys. Rev. B {\bf 27}, 3494 (1983).


\end{references}

\end{multicols}
\end{document}